# Morphological dilation image coding with context weights prediction

Jiaji Wu[1,2], Yan Xing[1], Anand Paul[2], Yong Fang[3], Jechang Jeong[2], Licheng Jiao[1], and Guangming Shi[1]

[1]Key Laboratory of Intelligent Perception and Image Understanding of Ministry of Education of China, Institute of Intelligent Information Processing, Xidian University, Xi'an 710071, China

[2]Dept. Electronics and Computer Engineering, Hanyang University, Seoul 133-791, Korea

[3]College of Information Engineering, Northwest A&F University, Yangling, 712100, China

**Abstract**: This paper proposes an adaptive morphological dilation image coding with context weights prediction. The new dilation method is not to use fixed models, but to decide whether a coefficient needs to be dilated or not according to the coefficient's predicted significance degree. It includes two key dilation technologies: 1) controlling dilation process with context weights to reduce the output of insignificant coefficients, and 2) using variable-length group test coding with context weights to adjust the coding order and cost as few bits as possible to present the events with large probability. Moreover, we also propose a novel context weight strategy to predict coefficient's significance degree more accurately, which serves for two dilation technologies. Experimental results show that our proposed method outperforms the state of the art image coding algorithms available today.

**Keywords:** quad-tree coding, morphological dilation, variable-length group test coding, weights training.

# 1. Introduction

Several very competitive wavelet-based image compression algorithms have been developed in past more than ten years. One kind of them is spatial tree-based including Embedded Zerotree Wavelet (EZW) algorithm [1] which was presented by Shapiro in 1993 , and Said and Pearlman's Set Partitioning In Hierarchical Trees (SPIHT) algorithm [2] presented in 1996 and Danyali's highly scalable image compression based on SPIHT for network applications presented in 2002 [3]. Bayazit proposed an enhanced SPIHT image coding algorithm using R-D optimization and high-order context arithmetic coding [4]. This kind of zerotree methods exploits magnitude correlation cross-subband of the decomposition, but it largely ignores the within-subband correlation. Thus another kind of algorithms which are block-based, such as Set Partition Embedded bloCK (SPECK) [5], Embedded ZeroBlock Coding and context modeling (EZBC) [6, 7] and JPEG2000 [8] et al, is applied to wavelet image coding. Compared with zerotree coding, the main strength of block-based coding algorithms is that it can more efficiently reduce the correlation of within-subband. However, block-based coding algorithms restrict searching significant coefficients inside block boundaries because they employ zeroblock coding based on quad-tree decomposition. In 2006, wavelet based embedded image coding using unified zero-block-zerotree approach [9] was proposed. The combination of zero-block and zero-block can break the searching restrict of block boundary. Reference [9] shows the performance of the method is higher than SPIHT and SPECK.

Morphological dilation coding algorithms, such as Servetto *et al.'s* morphological representation of wavelet data (MRWD) [10] and Chai *et al.'s* significance-linked connected component analysis (SLCCA) [11], also exploit within-subband correlation. MRWD exploits within-subband clustering of significant coefficients using conditioned dilation operation to search and code significant coefficients. SLCCA strengthens MRWD by exploiting not only within-subband clustering of significant coefficients but also cross-subband dependency in the significant fields. The cross-subband dependency is effectively exploited by using parent-child relationship. However, the drawback of dilation algorithms is that the seed of a significant cluster cannot be found quickly by zig-zag scan because the energy of a transformed image is not distributed in a regular shape.

Through analyzing the disadvantages of quad-tree coding and present dilation methods, a direct and perceptively improved coding method is to integrate morphological dilation operation with quad-tree coding [12]. However, many drawbacks of present dilation methods limit the performance of the whole coding method. First, no matter how many significant coefficients among the $N$ available coefficients which need to be dilated around a seed, the ordinary dilation methods always need $N$ bits to represent these significant or insignificant coefficients. In other words, in the conventional morphological dilation algorithms, even though the coefficient preparing to be dilated is insignificant, the codec also has to cost 1 bit to present it. For $N$ dilated coefficients which are regard as a group, each of them needs 1 bit to present it on matter whether the coefficient is significant or not.

Let $M$ denote the number of significant coefficients among the $N$ available coefficients in bitplane coding. For transformed image, because of its sparse representation, the appearance probabilities of different $M$ are not equal in subbands LH, HL, and HH but decrease as $M$ increases. The energy of the transformed image is mainly focused on a few of large coefficients, and most of coefficients are very small. Therefore, in most of cases, the smaller $M$ has higher probability to appear. In conventional dilation coding methods, using the fixed average codeword length is obviously not perfect. In addition, according to dilation template, traditional dilation methods [10, 11] dilate the $N$ available coefficients around a significant seed one by one with fixed order, and hardly consider the significance degrees of these coefficients. In fact, we can dilate these coefficients according to their significance degrees, only dilate those coefficients with higher significance degree and neglect those with lower significance degree.

To overcome the drawbacks of the ordinary dilation methods, in our paper, we propose a new morphological dilation coding method which includes two key technologies. One is to control dilation process by context weights; the other is variable-length group test coding method with context weights. The first technology is only to dilate some coefficients whose predicted significance degrees are large by using trained weights. The second technology can be divided into two facets, variable-length group test coding is to represent the events with large probability with as few bits as possible, while context weights are to adjust the order of dilation so that the probability of those events needing fewer bits becomes larger. It is well known that the context is applied widely in high-order entropy coding, but the paper is only concerned with bitplane coding of wavelet coefficients by using context not entropy coding.

The rest of this paper is organized as follows: section 2 introduces the most simple morphological dilation image coding with quad-tree partitioning. Section 3 presents our new algorithm which studies the method of training weights and two dilation strategies. Experimental results are reported in section 4 and some concluding remarks are made in section 5.

## 2. Morphological dilation coding with quad-tree partitioning

In our paper, we have analyzed the disadvantages of the quad-tree and dilation coding methods. Quad-tree partitioning only uses within-subband correlation and cannot break the limit of block boundary. Morphological dilation method can exploit cross-subband dependency, but it cannot quickly find a seed of dilation by zig-zag scan. According to these drawbacks, it is found that integrating quad-tree coding with morphological dilation method is undoubtedly logical. They can overcome the disadvantage of each other. In the following, we introduce a simple morphological dilation image coding framework with quad-tree partitioning, because this simple method is the foundation of our coding framework.

In the dilation coding framework, quad-tree partitioning is used to find a significant coefficient regarded as seed. We assume that a given block set S is processed by testing it for significance against the bitplane threshold. If S is not significant, it stays in the LIS (List of Insignificant Sets) and a "0" bit is outputted. If significant, output a "1" bit, then S is partitioned into four subsets, each of which has size approximately one-fourth the size of the parent set S. In the following procedure, each of these offspring sets is tested for significance for the same bitplane threshold, if significant, quadrisect once more. If not significant, it is added to the LIS. Each significant subset is, in turn, treated as a set of type S and processed recursively, until pixel-level is reached. When a significant pixel in the original set S is detected, then we consider it as seed to start dilating. The detail of quad-tree partitioning is introduced in [5].

Dilation operation is used on every newly found seed to find significant cluster. When the whole significant cluster is found, quad-tree partitioning is resumed in order to find next seed. The dilation and quad-tree coding perform recursively until the whole subband is exhausted. Most dilation methods use the dilation templates shown in Fig.1 (a) or Fig.1 (b).

Fig.2 is a demonstration of the progressive cluster detection by using our coding framework on a simple example. The 4-connected structuring element (dilation templet) shown in Fig.1 (a) is used here. Dilation operation produces an enlarged set containing the seed and some neighboring pixels by using the structuring element. In this demonstration, a significant seed coefficient can be quickly found by using quad-tree partitioning, then, the rest coefficients in whole cluster can be found by dilation operation.

However, as we analyze in section 1, one drawback of present dilation method is to use nearly fixed $N$ bits to represent the $N$ dilated coefficients, regardless of the number of significant coefficients among *N* dilated coefficients.

Another drawback is that the coefficients always are encoded one by one and "first dilation first coding" with fixed order. In the case, codec regards all coefficients have the same degree of significance. In fact, each coefficient maybe has different significance degree. If we can code the coefficients with large degree of significance earlier, and neglect those with low degree of significance, a more efficient bitplane coding could be design.

In the following sections, we will introduce our proposed algorithms to optimize the coding process.

## 3. Optimizing morphological dilation coding by context weights

In dilation process of our proposed algorithm, we employ the dilation model shown in Fig.1 (b), which includes eight neighbors. Due to the disadvantages of simple dilation methods and sparse representation of transformed images, clearly, in subbands LH, HL, HH, there will output lots of "0" bits to represent insignificant coefficients around a seed for the model shown in Fig.1 (b). If we could avoid or reduce these "0" bits, the coding performance would be improved.

In order to achieve this purpose, in subbands LH, HL, HH, we adopt two effective dilation strategies: one is controlling dilation process by context weights, the other is using variable-length group test coding method with context weights. Before introducing these two strategies, we propose weights training to predict significance degrees of coefficients which will be used in both of the two strategies.

### 3.1 Predicting significance degrees of coefficients

Although wavelet transform is applied to remove the vertical and horizontal linear redundancy, there still exist various correlations between the neighboring transformed coefficients. It mainly contains within-subband correlation such as zero clusters and sign correlation, and cross-subband correlation such as tree structure and parent-child relationship. Thus, the significance degree of the current coefficient can be estimated by its neighbors' significances according to the within-subband correlation and cross-subband correlation.

In this paper, we propose a novel method to predict significance degrees of coefficients with least square method.

First, in view of the within-subband correlation and cross-subband correlation in transformed image, the predicted value $x'_{i,j}$ of the current coefficient $x_{i,j}$ can be obtained by the model in Fig.3:

$$x'_{i,j} = \sum_{i=1}^{12} \alpha_i B_i + \alpha_{13} B_p + \alpha_{14} B_c. \tag{1}$$

Here, $B_1 \sim B_{12}$ are spatially adjacent coefficients of $x_{i,j}$. $B_p$ is parent coefficient of $x_{i,j}$. $B_c$ is the average value of four children coefficients of $x_{i,j}$. $\alpha_i (i = 0,1,2,\cdots,14)$ are so-called weights, whose values respectively indicate the effect degrees of the corresponding neighbors on the current coefficient.

Second, assuming training the weights $\alpha_i (i = 0,1,2,\cdots,14)$ in a local slip window which includes $L \times L$ coefficients $x_{1,1}, x_{1,2},\cdots,x_{L,L}$, we can get the estimated value of every coefficient in the window by (1). Thus,

$$\vec{x}' = C\vec{\alpha}, \tag{2}$$

where $\vec{x}' = [x'_{1,1}, x'_{1,2},\cdots,x'_{L,L}]^T$ is the vector of all the coefficients' estimated values, $\vec{\alpha} = [\alpha_1, \alpha_2,...,\alpha_{13},\alpha_{14}]^T$ is the vector of all the weights, and $C$ is a data matrix whose $Kth$ row vector is the fourteen nearest neighbors of the $Kth$ $x$.

Third, the minimum of mean square error is:

$$\min \left\| \vec{x} - \vec{x}' \right\| \text{ or } \min \left\| \vec{x} - C\vec{\alpha} \right\|. \tag{3}$$

Here, $\vec{x} = [x_{1,1}, x_{1,2}, \cdots, x_{L,L}]^T$ is the vector of all the coefficients' values. Evaluate the minimum of mean square error, we can get weight vector $\vec{\alpha}$:

$$\vec{\alpha} = (C^T C)^{-1} C^T \vec{x}. \tag{4}$$

Fourth, there is direct relation between the value of the current coefficient and its significance degree, and the trained weights respectively indicate the effect degrees of the corresponding neighbors on the current coefficient, so these weights can also reflect the contribution of the significance of the corresponding neighbors on the significance degree of the current coefficient. Thus the significance degree $W$ of coefficient can be estimated by the following equation:

$$W = \sum_{i=1}^{12} \alpha_i S_i + \alpha_{13} S_p + \alpha_{14} S_c. \tag{5}$$

Here, $S$ represents corresponding coefficient's significance, if the coefficient has been tested to be significant, $S = 1$; if the coefficient has been tested to be insignificant or hasn't been tested yet, $S = 0$.

The above four steps show the process of estimating significance degree of coefficient.

Here, it should be noticed that we can not directly get an optimal $W$ by (1) and (3). The reason is because $S_i$, $S_p$ and $S_c$ are binary values in (5). In this case, usually $C^T C$ in (4) is singular matrix, so we cannot obtain $C^T C$'s inversion by (4).

In addition, the weights $\alpha_i (i = 0, 1, 2, \cdots, 14)$ trained at encoder side need to be transmitted to the decoder side. If weights training are performed in smaller window, the weights will more approximate the local characteristic, but we will have to cost lots of bits to transmit these sets of weights. However, if we train these weights in a whole image, they cannot describe the details of the image. To take a tradeoff way, we train these weights in subbands HL, LH, HH respectively, (In subband LL, because nearly all of coefficients are significant, we only use the simple dilation method with quad-tree partition shown in section 2 and needn't to train weights.) and obtain three sets of weights of one image. These weights can not only cost fewer bits but also describe details of images very well. But in view of few coefficients are significant in the highest HH, LH, HL subbands, so the weight training excludes those highest frequency subbands.

With the significance degrees of coefficients predicted by the above method, we propose two effective strategies to control the dilation process and get better performance.

### 3.2 Control dilation process by context weights

In dilation process, some of the eight coefficients dilated from the same significant coefficient have been tested to be significant or insignificant already before this dilation process, which are not included in available coefficients. So the number of available coefficients dilated from the same one is different from 1 to 8, we use $N$ to represent this number, and it is known before coding. As our analysis for nature images shows, the energy of a transformed image is compacted in a few coefficients, the probability of finding an insignificant coefficient is much higher than that of finding a significant one, especially in subbands LH, HL, HH. So the number of significant coefficients in the group of all $N$ coefficients is always very small even zero, we use $M$ to represent the number of significant coefficients in the group, it varies from 0 to 8, while it is unknown before coding. For having a much better understanding, we also made a statistic shown in Fig.4 which describes that with the increasing of $M$, $M$'s appearance probability is quickly decreasing as a whole.

Here, we propose a dilation method to control dilation process by predicting significance degrees of coefficients. Before dilating the $N$ available coefficients around a seed, the codec calculates the value $W$ of every coefficient's significance degree first. The larger a coefficient's corresponding $W$, the greater the probability of the coefficient to be significant. The more significant a coefficient is, the earlier the coefficient must be dilated. Therefore, we dilate the $N$ available coefficients according to the values of their predicted significances degrees in descent order. Then we can stop outputting after dilating one insignificant coefficient, the remaining coefficients are most likely insignificant. In addition, from Fig.4, we can find that the probability of $M < 3$ is much larger than that of $M \geq 3$ in nature images. So when the first two coefficients dilated are all significant, we stop dilating, the rest may be insignificant in great proportion. But these remaining coefficients which are not dilated are not ignored, and they will be further tested to find a significant coefficient regarded as seed by quad-tree partitioning. Because the remaining coefficients are insignificant very likely, fewer bits are cost in the process of finding seed by quadtree. It is most possible that codec only needs a "0" bit to present lots of insignificant coefficients.

The main steps of this method are:

- find a seed by quad-tree partitioning;

- predict significance degrees of the $N$ available coefficients around the seed, then start dilating the $N$ coefficients according to the values of their predicted significance degrees $W$ in descent order;

- if the first two coefficients dilated are all significant, the codec stops dilating; if one insignificant coefficient is dilated, the codec stops dilating;

- with quad-tree partitioning, find a new seed to restart dilation in these coefficients which are not coded.

The method can nearly dilate all significant coefficients around a seed as $N$ is not very big. But when $N$ is bigger, the number of significant coefficients may be not smaller than 3, or the $W$ of an insignificant coefficient may be greater than that of a significant one. In this case, a few of significant coefficients will be not dilated, so we cannot find the whole significant cluster, and the bit output is not saved in finding a new seed by quad-tree partitioning. Therefore, in order to obtain better coding performance, for those cases that $N$ is larger, we propose a variable-length group test coding method with context weights to avoid outputting lots of "0" bits effectively and find the whole significant cluster quickly.

### 3.3 Variable-length group test coding method

In dilation process, the easiest way is to code these coefficients one by one, which is dilated from a same significant coefficient, and output a "1" bit for significant one and a "0" bit for insignificant one. No matter how many significant coefficients there are, it will output $N$ bits to code the $N$ available coefficients around a seed. For example, the output of this coding method with $N = 5$ is shown in Fig.5 (a).

While both the analysis for transformed nature images and Fig.4 show that only a few coefficients are significant, especially in subbands LH, HL, HH, the appearance probabilities of $M = 0$ and $M = 1$ are much bigger than those of $M = 2, 3, \cdots, 8$. This reminds us the quad-tree coding method which treats all the coefficients in a quad as a group. If there is at least one significant coefficient in the quad, codec outputs a "1" bit to represent the group is significant, and then partitions the quad into four subsets, each of which is tested separately; if there is no significant coefficient in the group, it is insignificant, codec just outputs a "0" bit to represent all these insignificant coefficients.

In fact, in dilation process, we can also use the quad-tree coding method to code these coefficients dilated from a same significant one, the only difference is that a group here does not always contain 4 coefficients but $N$ available coefficients, while this number is known before the dilation process. So, in dilation process, we treat all the coefficients dilated from a same significant one as a group, and test the group first. It is significant when there is at least one significant coefficient, codec outputs a "1" bit, and then tests these coefficients in the group separately; if there is no significant coefficient in the group, it is insignificant and codec outputs a "0" bit. We call this process group test. Fig.5 (b) shows the tree structure of group test when $N = 5$.

Take $N = 5$ for example, to do this, the output of $M = 0$ and $M = 5$ is "0" and "111111," so an event with big probability will just cost one bit while an event with small probability will cost six bits. The larger these big probabilities are, the fewer bits are used. Now, we prove the group test is better than coding directly.

If having $N$ available coefficients, for coding directly method, the average codeword length is $N$; for group test method, when $M = 0$, the average code length is 1, when $M > 0$, the average code length is little smaller than $N + 1$, so the average codeword length of the method is about

$$1 \times p_0 + (N+1) \times (1 - p_0), \tag{6}$$

where $p_0$ is the probability of $M = 0$ when the number of available coefficients is $N$. For natural images, in subbands LH, HL, HH, when $N$ is not very small, we can easily obtain:

$$p_0 > 1/N. \tag{7}$$

Thus we have
$$1 \times p_0 + (N+1) \times (1 - p_0) = N + 1 - p_0 N < N. \tag{8}$$

Equation (8) shows that the average codeword length of the group test method is smaller than that of coding directly.

From the above analysis mentioned and the output of direct coding shown in Fig.5 (a), it is easily to further think about that if we can know the exact value of $M$, then we can use it to control the coding process. In this case of knowing the value of $M$ in advance, once the number of significant coefficients equals to $M$ or the number of insignificant coefficients equals to $N-M$, we can stop coding immediately, and the following coefficients must be insignificant or significant without any test and output.

In the light of this, we can save lots of bits in our coding process especially in the case of $M=1$. Thus we propose a new group test method by using variable length group test coding. The main idea of the method is:

- for $M=0$, the group is insignificant, we use one "0" bit to represent;
- for $M>0$, the group is significant, we use one "1" bit to represent, then use another pre-bit before coding these coefficients to represent $M$, output a "1" bit for $M=1$, else a "0" bit for $M>1$;
- In the following coding process, for $M=1$, once a "1" bit has been outputted, which means the only one significant coefficient has been dilated, the codec stops coding, because the following coefficients must be insignificant without test; for $M>1$, there are at least two significant coefficients, once $N-2$ "0" bits have been outputted, which means $N-2$ insignificant coefficients have been dilated, then the codec stops coding, the following coefficients, if there are any, must be significant without any test and output.

In the new method, as Fig.6 shows, to code groups with the same number coefficients maybe cost different number of bits, so we name it variable-length group test coding method.

For $M=0$, only one bit is needed, thus its average codeword length of the group is 1; for $M=1$, the average codeword length is $2+(1+2+\cdots+N-2+N-1+N-1)/N$; for $M>1$, the average codeword length is little smaller than $2+N$, here we just use $2+N$ as the average codeword length when $M>1$. Then the average codeword length of the variable-length group test coding method is:

$$1\times p_0 +(2+(1+2+\cdots+N-2+N-1+N-1)/N)\times p_1 +(2+N)\times(1-p_0-p_1), \quad (9)$$

where $p_0$ is the probability of $M=0$, and $p_1$ is the probability of $M=1$ when the number of available coefficients is $N$.

Now we prove the variable-length group test coding method is superior to the original group test method which is not optimized by the variable-length output.

Similarly, for transformed natural images, in subbands LH, HL, HH, $p_0$ and $p_1$ are much larger than the probabilities of $M=2,3,\cdots,N$, so the following condition can be satisfied easily as $N$ is not very small:

$$\min(p_0, p_1) > 2N/(N^2+N+2). \quad (10)$$

The difference between the original group test method shown in Fig.5 (b) and variable-length group test coding method in average codeword length is:

$$\begin{aligned}&[1\times p_0 +(2+(1+2+\cdots+N-2+N-1+N-1)/N)\times p_1 +(2+N)\times(1-p_0-p_1)] \\ &-[1\times p_0 +(N+1)\times(1-p_0)] \\ &=1-p_0 -(\frac{N^2-N+2}{2N})p_1 \\ &<1-(\frac{N^2+N+2}{2N})\min(p_0,p_1) .\end{aligned} \quad (11)$$

With the condition of (10), we can obtain (11) is smaller than zero. So

$$\begin{aligned}&1\times p_0 +(2+(1+2+\cdots+N-2+N-1+N-1)/N)\times p_1 +(2+N)\times(1-p_0-p_1) \\ &<1\times p_0 +(N+1)\times(1-p_0) .\end{aligned} \quad (12)$$

Equation (12) shows the variable-length group test coding method is better than the original group test. Although we use one more bit to represent $M$, more bits are saved in the following coding process. So the one pre-bit is worthy to use.

In Fig.5 and Fig.6, we take $N=5$ as example and compare the outputs of the three different methods including coding directly, group test coding, and variable-length group test coding.

### 3.4 How to choose dilation strategy?

In section 3.2 and 3.3, we have discussed two dilation strategies — controlling dilation process by context weights and using variable-length group test coding method, respectively. It is found that only when the dilation strategies satisfy some conditions can they be used effectively. We analyze the condition of using variable-length group test coding method firstly.

The condition of variable-length group test coding over coding directly is:

$$[1 \times p_0 + (2 + (1 + 2 + \cdots + N - 2 + N - 1 + N - 1)/N) \times p_1 + (2 + N) \times (1 - p_0 - p_1)] - N$$
$$= 2 - (1 + N) p_0 - (\frac{N^2 - N + 2}{2N}) p_1 < 0 \quad . \quad (13)$$

Because

$$2 - (1 + N) p_0 - (\frac{N^2 - N + 2}{2N}) p_1 \leq 2 - (\frac{3N^2 + N + 2}{2N}) \min(p_0, p_1), \quad (14)$$

the condition of (13) can be converted into:

$$2 - (\frac{3N^2 + N + 2}{2N}) \min(p_0, p_1) < 0 \quad or \quad \min(p_0, p_1) > \frac{4N}{3N^2 + N + 2}. \quad (15)$$

Thus, if (15) can be satisfied, we can just use variable-length coding method. However, what regretful is that the value of $p_0$ and $p_1$ is unknown before coding. This leads (15) not to be judged. In view of the probabilities $p'_0$ and $p'_1$ of $M = 0$ and $M = 1$ under the case that the number of available coefficients is $N$ in the coded coefficients are near to $p_0$ and $p_1$, and the probabilities $p'_0$ and $p'_1$ is known, we use $p'_0$ and $p'_1$ to judge variable-length coding method can be used or not instead of $p_0$ and $p_1$.

$$\min(p'_{0,} p'_{1}) > \frac{4N}{3N^2 + N + 2}. \tag{16}$$

It is noted that $N$ is generally larger when (16) is satisfied. That is, variable-length group test coding method is not suitable to the case that $N$ is smaller. However, according to the description in section 3.2, controlling dilation process by context weights is more suitable to the case that $N$ is smaller. So we use the method that controlling dilation process by context weights instead of variable-length group test coding method when (16) is not satisfied.

### 3.5 Optimize variable-length group test coding using context weights

Let us focus on Fig.6 once more, at the bottom of the tree structure, there are some different cases respectively in every branch. Generally, the probabilities of these cases are equal, but the length of bits used to code each case is different. How many bits it will cost is not random, but is regular. It depends on the position of the significant coefficients. When $M = 1$, if one significant coefficient has been tested, we can stop dilating, the remnant coefficients must be insignificant without test. Therefore, the earlier the only significant coefficient is coded, the fewer bits will be needed. When $M > 1$, if $N - 2$ insignificant coefficients have been dilated and coded, the codec stops coding, the following coefficients must be significant. So the later the significant coefficients are coded, the fewer bits will be needed. However, how can we adjust the coding (or dilation) order as the way shown in Fig.7 and make the significant coefficients coded earlier or later?

We resume the method of predicting coefficient's significance degree which is presented in section 3.1 to solve this problem. Although the coefficient's significance degree predicted by the coefficient's contexts can not represent whether the coefficient is significant or not exactly, at a certain extent, it can represent the probability of being a significant coefficient. Thus we can change the coding order according to the significance degrees of the untested group coefficients.

The main procedure is:

- before dilating and coding the untested coefficients of a group, predict their significance degrees by (5), and then dilate the coefficients according to the values of their significance degrees;

- when $M=1$, for the significant coefficient to be coded earlier, we arrange the coding order according to the estimated significance degrees in descent way; for $M>1$, for the significant coefficients to be coded later, we arrange the coding order according to the estimated significance degrees in ascend way.

If these coefficients are rearranged in perfect order, we only need one bit in $M=1$ and $N-2$ bits in $M=2$ which are much less than the bits needed in variable-length group test coding method without context weights.

The average codeword length of variable-length group test coding method with context weights is:

$$1\times p_0 + (2+1)\times p_1 + (2+(N-2))\times p_2 + (2+N)\times(1-p_0-p_1-p_2). \qquad (17)$$

Although we can't make the coding order in such perfect way, we can also rearrange the coding order into a better way, in other words, the probabilities of the events using fewer bits can be greatly enhanced. So the method can effectively reduce the average codeword length of a group.

### 3.6 Complete coding procedure

In the above chapters, we proposed two dilation technologies for subbands LH, HL, HH. The first is to control dilation process by context weights, which is more suitable to the case that $N$ is smaller. The second is variable-length group test coding with context weights, which is effective when (16) is satisfied. In the case that (16) can not be satisfied, $N$ is generally smaller, and so controlling dilation process by context weights is more effective in this case. Thus, in order to obtain better coding performance, we combine the two strategies, use the method that controlling dilation process by context weights when (16) is not satisfied, and use the variable-length group test coding with context weights when (16) is satisfied.

In addition, when coding subband LL, we do not use our dilation method but use the simple dilation method with quad-tree partition shown in section 2, because nearly all of coefficients in subband LL are significant.

We give the whole coding process in Fig.8.

## 4. Experiments and results

Our experiments are performed on the $512 \times 512$, 8-bit standard gray images including Barbara, Lena, Goldhill, Baboon, Tank and Finger using 5-level 9/7 floating Daubechies filers [12]. The new method without arithmetic coding is compared with SPECK [5], SPIHT [2], EZBC [7], Modified SPIHT [4] and WBTC [9]. The PSNR results for SPECK-AC [5], SPIHT-AC [2] and JPEG2000 [8] and the new method with arithmetic coding are also provided. In our proposed method with arithmetic coding, significance coding, sign coding, and refinement coding use different independent zeroth-order arithmetic coding respectively. In addition, we give the comparison of SLCCA [11] and our proposed method.

Table 1 shows the PSNR results of some popular state-of-art image coding methods at the rates 0.125, 0.25 0.5, 1, and 2 bit per pixel (bpp). Without arithmetic coding, our new algorithm consistently outperforms SPECK and SPIHT. Compared with SPECK, the new algorithm is superior by 0.22dB on average. When compared with SPIHT, the new algorithm gains 0.33dB on average.

In order to further verify the performance of our new algorithm, we also compare the PSNR of several coding methods with arithmetic coding in Table 1. Table 2 also shows the PSNR performances of EZBC, Modified SPIHT, WBTC and our method without arithmetic coding in binary coding. Judging from this, our algorithm without arithmetic coding is better than the state of the art image coding methods obviously. In arithmetic coding, our new algorithm outperforms SPECK-AC, SPIHT-AC and JPEG2000 as well. Performances on Lena image show that the new algorithm is superior to SPECK-AC by 0.42dB, to SPIHT-AC by 0.18dB and also outperforms JPEG2000 by 0.24dB on average. Compared with SPECK-AC, SPIHT-AC and JPEG2000, on Goldhill image, the new algorithm gets 0.33dB, 0.16dB, and 0.14dB gains on average respectively, on Baboon image, the new algorithm gains 0.38dB, 0.15dB and 0.30dB on average respectively. PSNR performances on Barbara, Tank and Finger image also show that the new algorithm outperforms SPECK-AC, SPIHT-AC and JPEG2000 at most rates. The new algorithm is inferior to JPEG2000 only at 0.25bpp and 2bpp on Barbara image and to SPIHT-AC only at 2bpp on Lena image, whereas it is superior to SPECK-AC at all rates.

From table 1, we can observe that the new algorithm without arithmetic coding is still superior or comparable to SPECK-AC, SPIHT-AC, and JPEG2000 at most rates. According to this comparison, we can find the large dominance of our method in image coding.

Table 3 shows the PSNR comparison of SLCCA and our new algorithm. In present dilation methods, SLCCA is popular and excellent comparatively. However, compared to SLCCA, the new algorithm can get about 0.14dB gains on Lena, 0.13dB gains on Goldhill and 0.61dB gains on Tank on average. On Barbara, the new algorithm is slightly inferior to SLCCA at 0.125bpp and 0.25 bpp. Nevertheless, the new algorithm is superior to SLCCA as a whole.

From the above comparisons, it is found that the proposed method is not only superior to these block-based coding algorithms but also outperforms the classic dilation method ——— SLCCA. The large PSNR gain is mainly achieved from two aspects: 1) the method overcomes the disadvantage of block-based coding algorithms which restrict searched significant coefficients inside their block boundary; 2) weights are used to optimize the test order of the significant or insignificant coefficients in variable-length group test coding and obtain shorter average codeword.

Finally, let us compare the complexity of the proposed method, SPECK, SPIHT, JPEG2000 and SLCCA. At encoder side, the complexity of the proposed method is larger than SPECK and SPIHT because of on-line weights training. The proposed method obtains large PSNR gain at the cost of high complexity. Compared with JPEG2000 and SLCCA, although the proposed method needs to train weights on-line, it does not use complicated high-order context arithmetic coding. Considering synthetically, at encoder side, the complexity of the proposed method is nearly equivalent to JPEG2000 and SLCCA. However, at decoder side, the complexity of the proposed method is reduced in that the weights as side information are transmitted to the decoder side and need not to be trained. This makes the proposed method has equivalent complexity than SPIHT and SPECK, and lower complexity than JPEG2000 and SLCCA at decoder side. This asymmetric coding strategy owning high encoding complexity and reasonably low decoding complexity is more suitable to some applications which require high compression performance and low complexity at decoder side (rich server and slim client).

## 5. Conclusion

In our paper, we have proposed a novel morphological dilation image coding algorithm with context weights training. It combines quad-tree partitioning with two new dilation technologies effectively.

In view of the fact that the present dilation methods cannot quickly find a seed of dilation by zig-zag scan, we use quad-tree partitioning instead of zig-zag scan to search seed speedily.

Because of the sparse representation of transformed images, we mainly adopt two key technologies to optimize the process of dilation － controlling dilation process by context weights and using variable-length group test coding method. According to different cases, we employ different dilation strategies.

The experimental results have shown our method is a better coding method than the state of the art image coding algorithms available today. And also, without arithmetic coding, our method has very great potentiality.

To further improve performance, the high-order entropy coding can be studied by using the existing context template. In order to reduce the complexity of encoder side, we will also research efficient and robust training method offline to obtain weight prediction.

# List of figure captions

Fig.1. Different dilation templets: (a) includes 4 connected neighbors, and (b) includes 8 connected neighbors. Black and white blocks denote the seeks and corresponding dilation positions, respectively.

Fig.2. Demonstration of the progressive cluster detection by using simple dilation method with quad-tree partitioning on a simple example. Here, the dilation template shown in Fig.1 (a) is used. "□" denotes coefficient or block-set that is not encoded. "■" denotes significant coefficient found by quadtree partitioning. "■" denotes insignificant coefficient or block-set that is encoded by quadtree partitioning. "■" and "■" denote respectively significant and insignificant coefficients that is encoded by dilation.

Fig.3. 14th-order context model for training weights. $B_1 \sim B_{12}$ are spatially adjacent coefficients of $x_{i,j}$. $B_p$ is parent coefficient of $x_{i,j}$. $B_c$ is the average value of four children coefficients of $x_{i,j}$. Fig.4. Fourteenth-order context model for training weights. $x_{i-2,j}$, $x_{i-1,j-1}$, $x_{i-1,j}$, $x_{i-1,j+1}$, $x_{i,j-2}$, $x_{i,j-1}$, $x_{i,j+1}$, $x_{i,j+2}$, $x_{i+1,j-1}$, $x_{i+1,j}$, $x_{i+1,j+1}$, and $x_{i+2,j}$ are spatially adjacent coefficients of $x_{i,j}$. $x_p$ is parent coefficient of $x_{i,j}$. $x_c$ is the average value of four children coefficients of $x_{i,j}$

Fig.4. The probability of the number $M$ of significant coefficients in the group, here $M$ is from 0 to 8.

Fig.5. Tree structures of two coding methods when the number $N$ of available coefficients in a group is equal to 5. Black and white blocks denote significant and insignificant coefficients, respectively.

Fig.6. Tree structure of variable-length group test coding when the number $N$ of available coefficients in a group is equal to 5. Black and white blocks denote significant and insignificant coefficients, respectively.

Fig.7. Rearrange the coding order of coefficients to make the cases in left side become the cases in right side. Black and white blocks denote significant and insignificant coefficients, respectively.

Fig.8. The whole process of our coding method.

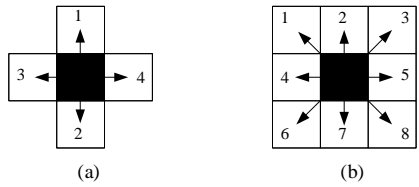

Fig.1. Different dilation templets: (a) includes 4 connected neighbors, and (b) includes 8 connected neighbors. Black and white blocks denote the seeks and corresponding dilation positions, respectively.

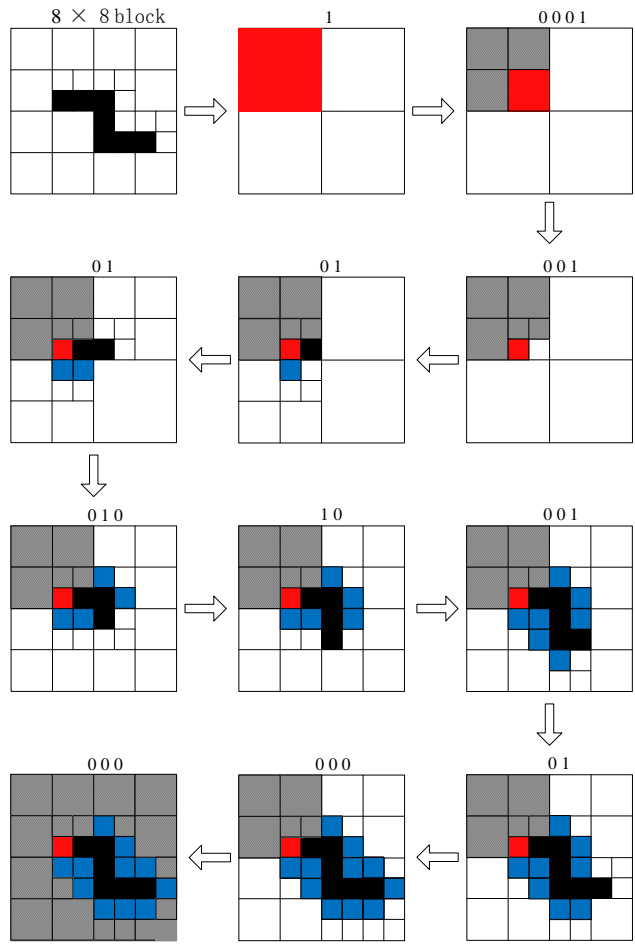

Fig.2. Demonstration of the progressive cluster detection by using simple dilation method with quad-tree partitioning on a simple example. Here, the dilation template shown in Fig.1 (a) is used. "☐" denotes coefficient or block-set that is not encoded. "🟥" denotes significant coefficient found by quadtree partitioning. "▨" denotes insignificant coefficient or block-set that is encoded by quadtree partitioning. "⬛" and "🟦" denote respectively significant and insignificant coefficients that is encoded by dilation.

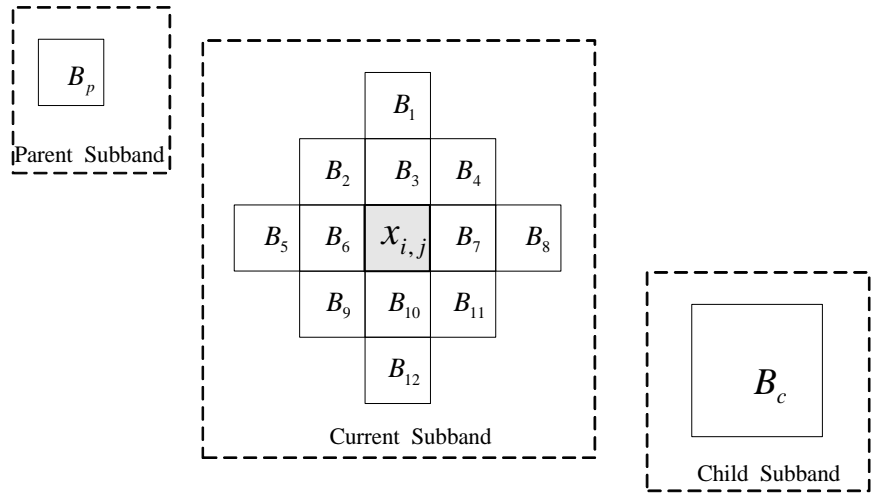

Fig.3. 14th-order context model for training weights. $B_1 \sim B_{12}$ are spatially adjacent coefficients of $x_{i,j}$. $B_p$ is parent coefficient of $x_{i,j}$. $B_c$ is the average value of four children coefficients of $x_{i,j}$.

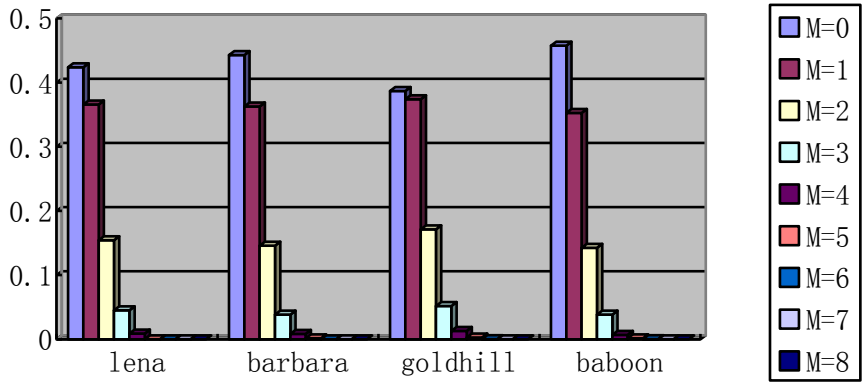

Fig.4. The probability of the number $M$ of significant coefficients in the group, here $M$ is from 0 to 8.

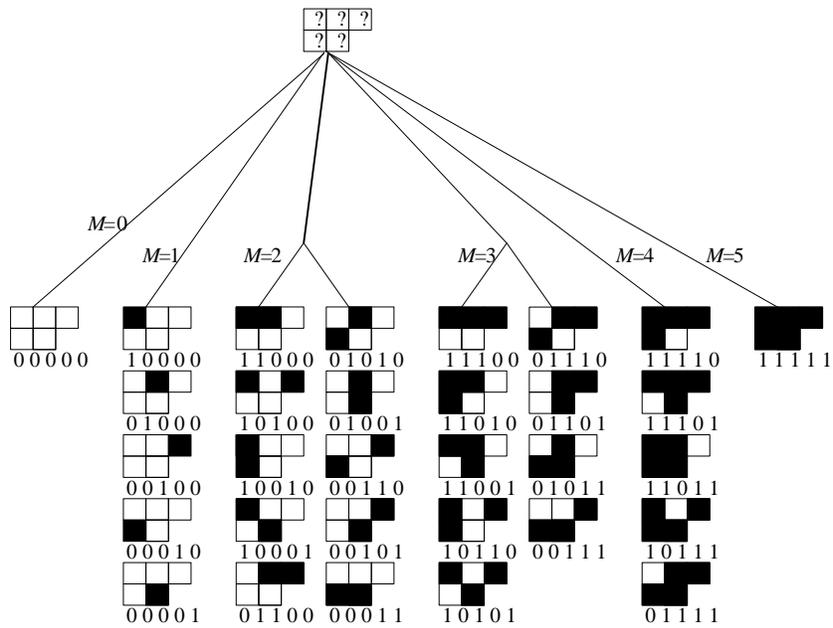

(a) The method of coding directly

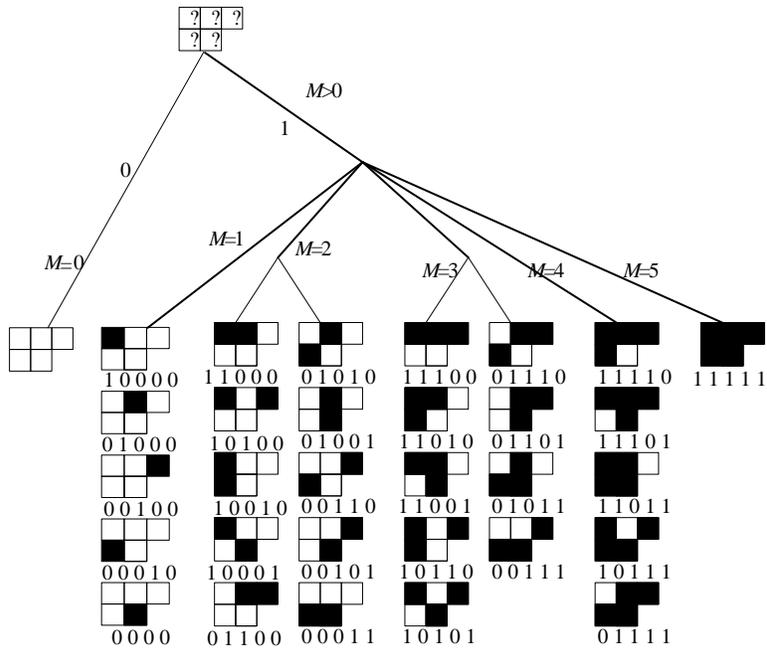

(b) The group test method

Fig.5. Tree structures of two coding methods when the number $N$ of available coefficients in a group is equal to 5. Black and white blocks denote significant and insignificant coefficients, respectively.

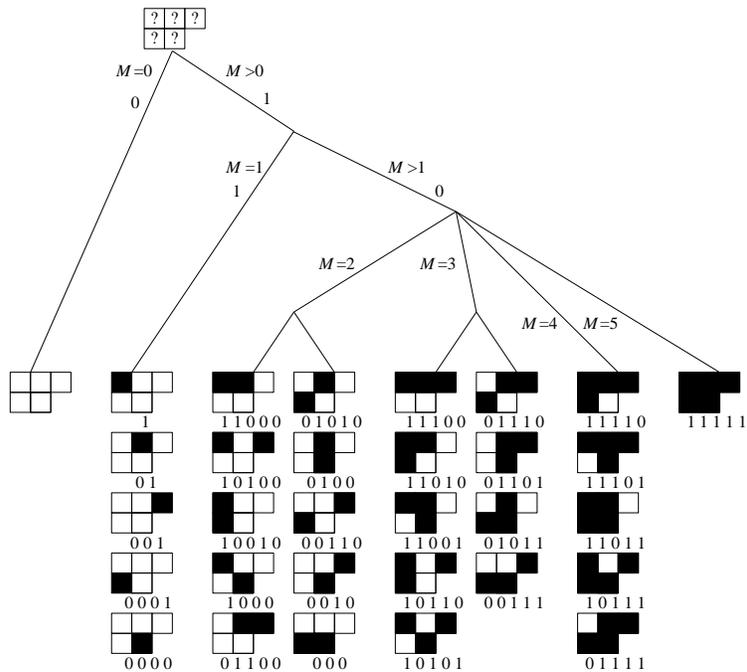

Fig.6 Tree structure of variable-length group test coding when the number *N* of available coefficients in a group is equal to 5. Black and white blocks denote significant and insignificant coefficients, respectively.

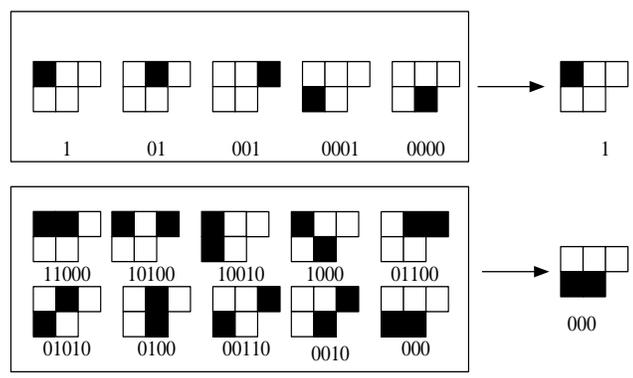

Fig.7. Rearrange the coding order of coefficients to make the cases in left side become the cases in right side. Black and white blocks denote significant and insignificant coefficients, respectively.

**The whole coding process**

1. Use wavelet transform for original image, and build context weights training model, then train three sets of weights in subbands HL, LH, HH.
2. Initialization.
● calculate $n_{max} = \lfloor \log_2(\max_{(i,j)\in Z}\{|c_{i,j}|\}) \rfloor$ as the highest bitplane;
● add the whole image to LIS, and set LSP=$\varnothing$;
● initialize 2D array P[8][2] with zero;
3. Start coding from the highest bitplane (MSB).
 Step 1: find a seed by quad-tree partition.
 Step 2: if the seed is in subband LL, use simple dilation method, output "1" for significant coefficient, and output a "0" bit for insignificant coefficient.
 Step 3: if the seed is not in subband LL,
  ■ get the number N of available coefficients around the seed, calculate the significance degrees $W$ of the N coefficients as Formula (5), judge Formula (16) is satisfied or not:
  ■ if Formula (16) is not satisfied, control the dilation process by context weights.
   ● arrange the coding order of the N coefficients according to their $W$ in descent way, then start coding;
   ● once an insignificant coefficient is dilated, stop dilating;
   ● once two significant coefficients are dilated, stop dilating;
   ● update the probabilities P[N][0] and P[N][1].
  ■ if satisfied, using variable-length group test coding with context weights.
   ● if M=0, output a "0" bit, update the probability P[N][0];
   ● else
   {
    -- output a "1" bit;
    -- if M=1,{
      output a "1" bit;
      arrange the coding order of the N coefficients according to their $W$ in descent way, then start coding;
      once a significant coefficient is dilated, stop coding;
      update the probability P[N][1].
     }
    -- else if M>1, {
      output a "0" bit;
      arrange the coding order of the N coefficients according to their $W$ in ascent way, then start coding;
      once N-2 insignificant coefficients are dilated, stop coding.
     }
   }
 Step 4: Goto Step 2 on every newly found significant coefficient, stop coding when the whole significant cluster is found.
 Step 5: Find a new seed of dilation by quad-tree partition again, repeat Step 2 or Step 3. The dilation and quad-tree coding perform recursively until the whole subband is exhausted.
 Step 6: Code the other subbands by Step 1~ Step 5.
 Step 7: Decrement current bitplane by 1, and start coding the next bitplane by Step 1 ~ Step 6.

Denote: N is the number of available coefficients in a group; M is the number of significant coefficients in a group. P[N][0] and P[N][1] denoting respectively the probabilities of M=0 and M=1 under the case that the number of available coefficients is N in the coded coefficients are equal to $p'_0$ and $p'_1$ in Eq. (16). During the whole encoding/decoding process, once current bit rate reaches the target bit rate, the process will stop.

Fig. 8. The whole process of our coding method.

# List of table captions

Table 1. PSNR (dB) performance comparison of different algorithms (SPECK, SPIHT, JPEG200, and our new method) at different bit rates.

Table 2. PSNR (dB) performance comparison of EZBC [7], Modified SPIHT [4], WBTC [9] and the proposed method without arithmetic coding.

Table 3. PSNR (dB) performance comparison of SLCCA and the proposed method with arithmetic coding at different bit rates.

**Table 1. PSNR (dB) performance comparison of different algorithms at different bit rates**

| Image | Bpp | Without Arithmetic Coding | | | With Arithmetic Coding | | | |
|---|---|---|---|---|---|---|---|---|
| | | SPECK | SPIHT | New | SPECK-AC | SPIHT-AC | JPEG2000 | New-AC |
| Barbara | 0.125 | 24.86 | 24.47 | **25.12** | 24.93 | 24.86 | 25.02 | **25.26** |
| | 0.25 | 27.62 | 27.22 | **28.07** | 27.76 | 27.58 | **28.27** | 28.13 |
| | 0.5 | 31.33 | 30.94 | **31.69** | 31.54 | 31.40 | 32.15 | **32.15** |
| | 1 | 36.27 | 35.94 | **36.55** | 36.49 | 36.41 | 37.11 | **37.21** |
| | 2 | 42.22 | 42.05 | **42.40** | 42.46 | 42.65 | **43.14** | 43.03 |
| Lena | 0.125 | 30.75 | 30.72 | **31.18** | 31.00 | 31.10 | 31.03 | **31.44** |
| | 0.25 | 33.76 | 33.70 | **34.15** | 34.03 | 34.12 | 34.15 | **34.42** |
| | 0.5 | 36.87 | 36.85 | **37.15** | 37.10 | 37.22 | 37.28 | **37.49** |
| | 1 | 40.02 | 39.99 | **40.17** | 40.25 | 40.42 | 40.35 | **40.57** |
| | 2 | 44.40 | 44.35 | **44.43** | 44.77 | **45.07** | 44.84 | 44.95 |
| Goldhill | 0.125 | 28.25 | 28.27 | **28.44** | 28.39 | 28.48 | 28.47 | **28.63** |
| | 0.25 | 30.25 | 30.22 | **30.40** | 30.50 | 30.56 | 30.54 | **30.69** |
| | 0.5 | 32.77 | 32.71 | **33.00** | 33.03 | 33.13 | 33.25 | **33.40** |
| | 1 | 36.08 | 36.00 | **36.31** | 36.36 | 36.55 | 36.60 | **36.83** |
| | 2 | 41.18 | 41.12 | **41.49** | 41.59 | 42.02 | 41.95 | **42.02** |
| Baboon | 0.125 | 21.55 | 21.49 | **21.60** | 21.63 | 21.72 | 21.50 | **21.75** |
| | 0.25 | 22.97 | 22.93 | **23.04** | 23.10 | 23.26 | 23.10 | **23.32** |
| | 0.5 | 25.27 | 25.21 | **25.42** | 25.40 | 25.64 | 25.52 | **25.80** |
| | 1 | 28.74 | 28.66 | **28.89** | 28.90 | 29.17 | 29.02 | **29.41** |
| | 2 | 34.26 | 34.15 | **34.50** | 34.60 | 34.98 | 34.83 | **35.23** |
| Tank | 0.125 | 27.92 | 27.90 | **28.01** | 28.06 | 28.11 | 28.04 | **28.21** |
| | 0.25 | 29.42 | 29.43 | **29.60** | 29.58 | 29.73 | 29.62 | **29.77** |
| | 0.5 | 31.45 | 31.44 | **31.67** | 31.73 | 31.82 | 31.83 | **31.97** |
| | 1 | 34.30 | 34.28 | **34.53** | 34.69 | 34.78 | 34.78 | **34.96** |
| | 2 | 38.95 | 38.99 | **39.14** | 39.67 | 39.81 | 39.69 | **39.95** |
| Finger | 0.125 | 21.78 | 21.65 | **21.92** | 21.93 | 21.87 | 21.73 | **22.05** |
| | 0.25 | 24.08 | 23.84 | **24.31** | 24.32 | 24.25 | 24.37 | **24.47** |
| | 0.5 | 27.40 | 27.17 | **27.61** | 27.79 | 27.67 | 27.86 | **28.02** |
| | 1 | 31.00 | 30.79 | **31.22** | 31.43 | 31.35 | 31.64 | **31.94** |
| | 2 | 36.48 | 36.35 | **36.78** | 37.26 | 37.14 | **37.60** | 37.56 |

**Table 2. PSNR (dB) performance comparison of EZBC [7], Modified SPIHT [4],
WBTC [9] and the proposed method without arithmetic coding**

| Images | bpp | EZBC | Modified SPIHT | WBTC | New |
|---|---|---|---|---|---|
| Lena (512x512) | 0.125 | 30.81 | 30.76 | 30.91 | **31.18** |
| | 0.25 | 33.80 | 33.73 | 33.82 | **34.15** |
| | 0.5 | 36.91 | 36.87 | 36.95 | **37.15** |
| | 1 | 40.06 | 40.02 | - | **40.17** |
| Goldhill (512x512) | 0.125 | 28.32 | 28.30 | 28.39 | **28.44** |
| | 0.25 | 30.29 | 30.23 | 30.29 | **30.40** |
| | 0.5 | 32.82 | 32.76 | 32.83 | **33.00** |
| | 1 | 36.16 | 36.10 | - | **36.31** |

Table 3. PSNR (dB) performance comparison of SLCCA and the proposed method with arithmetic coding at different bit rates

| Lena | | | | |
|---|---|---|---|---|
| bpp | 0.125 | 0.25 | 0.5 | 1 |
| SLCCA | 31.25 | 34.28 | 37.35 | 40.47 |
| NEW-AC | **31.44** | **34.42** | **37.49** | **40.57** |
| Goldhill | | | | |
| bpp | 0.125 | 0.25 | 0.5 | 1 |
| SLCCA | - | 30.60 | 33.26 | 36.66 |
| NEW-AC | **28.63** | **30.69** | **33.40** | **36.83** |
| Tank | | | | |
| bpp | 0.125 | 0.25 | 0.5 | 1 |
| SLCCA | - | 29.44 | 31.27 | 34.04 |
| NEW-AC | **28.20** | **29.77** | **31.97** | **34.96** |
| Barbara | | | | |
| bpp | 0.125 | 0.25 | 0.5 | 1 |
| SLCCA | **25.36** | **28.18** | 31.89 | 36.69 |
| NEW-AC | 25.26 | 28.13 | **32.15** | **37.21** |